\documentclass[12pt]{article}
\usepackage{graphicx}
\newcommand{\bb}{\begin{equation}}
\newcommand{\ee}{\end{equation}}
\newcommand{\ba}{\begin{eqnarray}}
\newcommand{\ea}{\end{eqnarray}}

\begin{document}

\title{{\bf Preserving the Sun from the Cold \\ by a Perfectly Reflecting Dyson Sphere}}

\author{
Don N. Page
\thanks{Internet address:
profdonpage@gmail.com}
\\
Department of Physics\\
4-183 CCIS\\
University of Alberta\\
Edmonton, Alberta T6G 2E1\\
Canada
}

\date{2022 July 21}

\maketitle
\large
\begin{abstract}
\baselineskip 20 pt

Some entities, such as humans, survive longest if their environment is neither too hot nor too cold, and the sun is no exception.  It is rather obvious that if the sun were enclosed inside a thermally conducting sphere surrounded by a heat bath kept much hotter than the present central temperature of the sun, its nuclear burning would occur faster, so that the sun would last for a shorter time.  It is less obvious that if the sun were surrounded by a perfectly reflecting sphere to prevent its radiation from escaping to cold empty space, it could actually last longer.  Here I shall show that this is the case for such a sphere at least somewhat larger than the present solar radius.  This na\"{\i}vely paradoxical result is a consequence of the negative specific heat of many gravitating systems, so as the energy emitted by the sun is reflected back to increase the thermal energy, the sun expands and its central temperature goes {\it down} rather than up and reduces the nuclear burning rate, so that the sun can last much longer than five billion years, for a lifetime growing roughly exponentially with the cube root of the radius of the perfectly reflecting sphere.  

\end{abstract}

\normalsize

\baselineskip 22 pt

\newpage

\section{Introduction}

A `Dyson sphere' \cite{Dyson} is a hypothetical construction by an advanced civilization around a star to collect far more of the stellar radiation than that intercepted by the civilization's planet (an idea attributed by Dyson \cite{DS,DS2} to Olaf Stapledon \cite{Stapledon}; another influence may have been J.\ D.\ Bernal \cite{Bernal}).  It has generally been thought of as a way for the civilization to utilize more of the stellar radiation, without considering the effect on the evolution of the star (which would generally be negligible in the more nearly realistic case in which the Dyson sphere intercepts only a small fraction of the stellar radiation).

Much of the research on Dyson spheres has been on their effect on the appearance of the system from far away, in order to provide a guide for potentially observational consequences for us.  Very little has been written on the effects on the central star.  The most recent papers that I am aware of on this are those of Huston and Wright \cite{H,HW}, which consider the effects on central stars from Dyson spheres that reflect just part of the stellar radiation back to the star.  Because of the negative gravitothermal heat capacity of stars, this influx of energy from the reflected radiation causes the star to expand and cool, reducing its luminosity, but generally only by less than a factor of two for the stars and partial reflections considered in those papers.

However, here I shall abandon all pretext of practicality and consider the thought experiment of what would happen if one did surround the sun by a perfectly reflecting spherical shell that did not let out any of the solar radiation.  Na\"{\i}vely one might expect preventing the radiation from escaping would cause the sun to heat up and increase the rate of nuclear burning, reducing its lifetime.  Nevertheless, the negative specific heat of gravitational systems can cause the sun's central temperature actually to cool as more solar energy is reflected back in, as well as swelling up the sun to reduce the central density, so that for any perfectly reflecting shell sufficiently larger than the present solar radius, the sun can burn slower and last much longer, all the way up to the timescale of baryon decay for shell radii comparable to 1 au.

\section{Solar Parameters}

The present sun is sufficiently condensed that its total gravitational energy, particle thermal energy, and radiation energy (this total being denoted by $E$, not including the rest mass energy of the particles) is too far negative for the sun to attain an isothermal configuration inside any perfectly reflecting sphere larger than the present solar radius $R_\odot$ before nuclear burning increases this energy $E$ sufficiently.  However, the time for the enclosed sun to increase $E$ sufficiently by nuclear burning so that an isothermal configuration can be achieved is comparable to the Ritter-Kelvin-Helmholtz timescale \cite{Giora} of about 16 million years (see below for the evaluation), which is much shorter than even the present calculated future lifetime of the sun (about 8 billion years, 5 during the Main Sequence \cite{Schroder:2008qn}), so I shall ignore this relatively short time and just estimate the time for the sun to burn its remaining hydrogen to helium when it is an enclosed isothermal plasma.  By then the temperature will be so hot that the remaining nuclear burning to iron will take much less time than the time of hydrogen burning, so I shall also ignore the time to burn helium to iron.

Once the sun has emitted enough energy that it becomes an isothermal ball of plasma, it will initially have negative specific heat, so that as $E$ increases, the temperature $T$ will decrease, down to a minimum determined by the mass $M_\odot$ of the sun, the radius $R$ of the perfectly reflecting sphere, and the mean mass $m$ of the particles in the sun.  Beyond this point, as $E$ continues to increase, the temperature will go back up.  The regime around the time of minimum $T$ will be a bottleneck for the nuclear burning, where the energy generation rate is lowest and most of the lifetime will be spent.  (Since the nuclear burning rate for fixed composition depends not only strongly on the temperature but also proportionally to the average square of the density over the configuration, which continually decreases as the energy $E$ increases and the self-gravitating plasma becomes less centrally condensed, the minimum nuclear burning rate occurs somewhat after the temperature reaches its minimum.)

To get the mean mass $m = \rho/n$ of the massive particles in the sun (e.g., not including the photons, but only the particles with nonrelativistic thermal energies), where $\rho$ is the mass density and $n$ is the total particle number density, for simplicity I shall neglect the mass of the electrons, assume a hydrogen nucleus mass fraction $X = 0.70$ (taken all to be protons), a helium nucleus mass fraction $Y = 0.28$ (taken all to be alpha particles), and mass fraction $Z = 0.02$ of `metals' (nuclei heavier than helium; here I shall take this small fraction to be oxygen-16) \cite{Schroder:2008qn}.  Approximating the mass of each baryon as the proton mass $m_p$ and neglecting the mass of the electrons, and setting the total number density to be $n = n_X + n_Y + n_Z + n_e$ with free proton number density $n_X$, alpha-particle number density $n_Y$, oxygen-16 nucleus number density $n_Z$, and electron number density $n_e \approx n_X + 2 n_Y + 8 n_Z$ for neutrality (the approximation coming from assuming that the metal nuclei all are those of oxygen-16, each with 8 protons), one gets $\rho_X \approx m_p n_X$, $\rho_Y \approx 4 m_p n_Y$, $\rho_Z \approx 16 m_p n_Z$, $\rho_e \approx 0$, and hence $\rho = \rho_X + \rho_Y + \rho_Z + \rho_e \approx m_p(n_X + 4 n_Y + 16 n_Z)$ and $n \approx 2 n_X + 3 n_Y + 9 n_Z$.  With $X = \rho_X/\rho \approx 0.70$, $Y = \rho_Y/\rho \approx 0.28$, and $Z = \rho_Z/\rho \approx 0.02$, this leads to
\bb
m = \frac{\rho}{n} \approx \frac{m_p}{2X+(3/4)Y+(9/16)Z} \approx \frac{m_p}{1.62125} \approx 1.03\times 10^{-27}\, \mathrm{kg}.
\label{m}
\ee

Now if along with this estimate for the mean particle mass $m$, we also use the solar parameters \cite{parameters} $M_\odot = 1.988\,41(4)\times 10^{30}$ kg, $GM_\odot = 1.327\,124\,400\,18(9)\times 10^{20}$ m$^3$ s$^{-2}$, nominal solar equatorial radius $R_\odot = 6.957\times 10^8$ m, nominal solar luminosity $L_\odot = 3.828\times 10^{26}$ W, and the Boltzmann constant $k \equiv 1.380\,649\times 10^{-23}$ J/K, we can calculate various other parameters for the sun, such as the total number of massive particles in the sun,
\bb
N_\odot = \frac{M_\odot}{m} \approx 1.93\times 10^{57},
\label{Nsun}
\ee
the average density of the sun,
\bb
\rho_\odot \equiv \frac{3M_\odot}{4\pi R_\odot^3} \approx 1\,410\ \mathrm{kg}/\mathrm{m}^3,
\label{rhosun}
\ee
the solar luminosity per mass
\bb
\epsilon_\odot \equiv \frac{L_\odot}{M_\odot} \approx 1.925\times 10^{-4}\ \mathrm{W}/\mathrm{kg} = 1.925\times 10^{-4}\ \mathrm{m}^2/\mathrm{s}^3
\label{epssun}
\ee
(which is about 1/2500 of an estimate of my own basal metabolic rate per mass and about 1/1000 of $c^2$ multiplied by the present Hubble expansion rate, so at its present luminosity the sun would emit about 0.1\% of its mass energy in a Hubble time),
a characteristic energy associated with the solar parameters,
\bb
E_\odot \equiv \frac{GM_\odot^2}{2R_\odot} = 1.897\times 10^{41}\ \mathrm{J},
\label{Esun}
\ee 
a characteristic solar temperature,
\bb
T_\odot \equiv \frac{E_\odot}{k N_\odot} \equiv \frac{GM_\odot m}{2kR_\odot} \approx 7.131\times 10^6\ \mathrm{K},
\label{Tsun}
\ee
and a characteristic solar time,
\bb
t_\odot \equiv \frac{E_\odot}{L_\odot} \equiv 
\frac{GM_\odot^2}{2R_\odot L_\odot} \approx 1.570\times 10^7\ \mathrm{years}.
\label{tsun}
\ee
This time is essentially the Ritter-Kelvin-Helmholtz timescale \cite{Giora}, the time for the sun to emit an amount of energy equal to the kinetic energy of its particles, which would also be an estimate for the future lifetime of the sun if no more nuclear reactions occurred.  The fact that the sun has lasted about 291 times as long, $4.568\times 10^9$ years \cite{Schroder:2008qn, Connelly}, is due to the nuclear burning of hydrogen into helium, which provides the heat and pressure to hold the sun up for billions of years against gravitational collapse.  If the sun were placed inside a perfectly reflecting sphere, it would take a time of the order of $t_\odot$ for the energy $E$ inside (not counting the decreasing rest mass energy of the particles as protons are converted to alpha particles, releasing the energy that gives the increase in $E$), which starts off negative because of the dominance of the negative gravitational potential energy, to become sufficiently close to zero (at a negative value depending on the radius of the reflecting shell, and on the mass inside, which I shall always approximate as $M_\odot$) that the self-gravitating plasma can become isothermal.  Because $t_\odot$ is so much smaller than the nuclear burning times, I shall ignore the time needed for the sun to evolve to an isothermal ball of plasma inside the reflecting sphere, though the details of that evolution could make another interesting research project, perhaps modeled on the papers of Macy Huston and Jason Wright cited earlier \cite{H,HW}.

\section{Properties of Self-Gravitating Spherically-Symmetric Isothermal Perfect Fluid Balls}

Now let us consider the properties of an self-gravitating isothermal ball of perfect fluid, such as the plasma of photons, electrons, protons, alpha particles, and oxygen nuclei that I shall take as an approximation for the solar composition (with oxygen standing in for all the nuclei heavier than alpha particles).  The ratio of the Schwarzschild radius of the sun to its present radius is
\bb
\frac{R_S}{R_\odot} = \frac{2GM_\odot}{c^2 R_\odot} = \frac{2\,953.250\,076\,50(20)\ \mathrm{m}}{695\,700\,000\ \mathrm{m}} \approx 4.245\times 10^{-6},
\label{RS/R}
\ee 
which is very small, so we can use the equations of nonrelativistic Newtonian gravitation, since I shall restrict to reflecting shell radii $R \geq R_\odot$.  The energy $E_r = (4\pi/3)R^3 a T^4$ of thermal electromagnetic radiation, with radiation energy density constant $a = \pi^2 k^4/[15(\hbar c)^3] \approx 7.566\times 10^{-16}$ J/(m$^3$K$^4$), can make a significant contribution to the total energy $E$ inside the shell that also includes the thermal energy of the nonrelativistic particles and the negative gravitational potential energy (but not the rest mass energies), so I shall also include this radiation energy.  However, I shall ignore energy losses to neutrinos.

The radial structure of a nonrelativistic self-gravitating spherically symmetric ball of perfect fluid is given in dimensionless form by the Emden-Chandrasekhar equation \cite{E,C}
\bb
\frac{1}{\xi^2}\frac{d}{d\xi}\left(\xi^2\frac{d\psi}{d\xi}\right) = e^{-\psi},
\label{EC}
\ee
where
\bb
\xi = r/L
\label{xi}
\ee
is a dimensionless radial variable, the physical radius $r$ (which ranges from 0 to the radius $R$ of the perfectly reflecting shell) divided by the length scale
(which should not be confused with the solar luminosity $L_\odot$)
\bb
L = \left(\frac{kT}{4\pi G m \rho_c}\right)^{1/2},
\label{L}
\ee 
where $T$ is the (constant in space) temperature, $m$ is the mean particle mass, and $\rho_c = \rho(r=0)$ is the central density of the fluid.
Furthermore,
\bb
\psi(\xi) = \ln{[\rho_c/\rho(\xi)]}
\label{psi}
\ee
is a logarithmic dimensionless measure of the inverse density as a function of the dimensionless radial variable $\xi$, so $\rho = \rho_c e^{-\psi}$.

The boundary conditions at the center, $\xi = 0$, with no point mass present there but just a smooth fluid density configuration, are that there $\psi = 0$ and $d\psi/d\xi = 0$.  This leads to a unique solution for $\psi(\xi)$, which one can readily calculate to have the following series form for small $\xi$:
\bb
\psi(\xi) = \frac{1}{6}\xi^2-\frac{1}{120}\xi^4+\frac{1}{1\,890}\xi^6-\frac{61}{1\,632\,960}\xi^8+\frac{629}{224\,532\,000}\xi^{10}+O(\xi^{12}).
\label{psiseries}
\ee
Series for other useful functions of $\xi$ and $\psi$, using $\psi'$ for $d\psi/d\xi$, are
\bb
\frac{\rho}{\rho_c}=e^{-\psi}=1-\frac{1}{6}\xi^2+\frac{1}{45}\xi^4-\frac{61}{22\,680}\xi^6+\frac{629}{2\,041\,200}\xi^8+O(\xi^{10}),
\label{rhoseries}
\ee
\bb
\frac{\rho_c}{\rho}=e^{\psi}=1+\frac{1}{6}\xi^2+\frac{1}{180}\xi^4-\frac{1}{11\,340}\xi^6+\frac{1}{510\,300}\xi^8+O(\xi^{10}),
\label{erhoseries}
\ee
\bb
\psi' \equiv \frac{d\psi}{d\xi} = \frac{1}{3}\xi-\frac{1}{30}\xi^3+\frac{1}{315}\xi^5-\frac{61}{204\,120}\xi^7+\frac{629}{22\,453\,200}\xi^9+O(\xi^{11}),
\label{psi'series}
\ee
\bb
u \equiv \frac{\xi e^{-\psi}}{\psi'}=3-\frac{1}{5}\xi^2+\frac{19}{1\,050}\xi^4-\frac{118}{70\,875}\xi^6+\frac{33\,661}{218\,295\,000}\xi^8+O(\xi^{10}),
\label{useries}
\ee
\bb
v \equiv \xi\psi' = \frac{1}{3}\xi^2-\frac{1}{30}\xi^4+\frac{1}{315}\xi^6-\frac{61}{204\,120}\xi^8+\frac{629}{22\,453\,200}\xi^{10}+O(\xi^{12}),
\label{vseries}
\ee
\bb
w \equiv uv \equiv \xi^2e^{-\psi} = \xi^2-\frac{1}{6}\xi^4+\frac{1}{45}\xi^6-\frac{61}{22\,680}\xi^8+\frac{629}{2\,041\,200}\xi^{10}+O(\xi^{12}),
\label{wseries}
\ee
\bb
\frac{\rho_c}{\bar{\rho}}=\frac{\xi^2}{3v}=1+\frac{1}{10}\xi^2+\frac{1}{2\,100}\xi^4-\frac{1}{121\,500}\xi^6+\frac{83}{327\,442\,500}\xi^8+O(\xi^{10}),
\label{rho_c/bar{rho}}
\ee
where $\bar{\rho}$ is the mean density inside $r = L\xi$.  At $\xi=1$, the five terms given for the last series are larger than $\rho_c/\bar{\rho}$ at this $\xi$ by only about one part in a million.

The derivatives of $u$, $v$, and $w$ can be written in terms of these same quantities as follows:
\bb
\xi\frac{du}{d\xi}= r\frac{du}{dr} = u(3-u-v),
\label{du}
\ee
\bb
\xi\frac{dv}{d\xi}= r\frac{dv}{dr} = v(u-1),
\label{dv}
\ee
\bb
\xi\frac{dw}{d\xi}= r\frac{dw}{dr} = w(2-v).
\label{dw}
\ee

Therefore, one can eliminate $\xi$ from the Emden-Chandrasekhar Eq.\ (\ref{EC}) \cite{E,C} to write it as an autonomous first-order equation between $u$ and $v$, $u$ and $w$, or $v$ and $w$:
\bb
v(u-1)du = u(3-u-v)dv,
\label{dudv}
\ee
\bb
w(2u-w)du = u(3u-u^2-w)dw,
\label{dudw}
\ee
or
\bb
w(2-v)dv = (w-v)dw.
\label{dvdw}
\ee

One can see that as $\xi$ becomes large, $u$ oscillates around 1, whereas $v$ and $w$ oscillate around 2.  Indeed, asymptotically
\bb
\frac{\rho}{\rho_c} = e^{-\psi} = \frac{2}{\xi^2}\left[1+\frac{a}{\sqrt{\xi}}\cos{\left(\sqrt{7/4}\ln{\xi}+b\right)}+O\left(\frac{1}{\xi}\right)\right]
\label{rhoasymp}
\ee
with constants $a$ and $b$ that can be determined by numerically integrating the Emden-Chandrasekhar Eq.\ (\ref{EC}) \cite{E,C} but which will not be important to us here.  One can see that, leaving aside the decaying oscillatory term, the density $\rho$ asymptotically decreases inversely proportional to the square of the radial distance $r = L\xi$, so the mass inside $r$ asymptotically increases linearly with $r$ and diverges as $r$ is taken to infinity.  Therefore, an isothermal self-gravitating fluid with no outer boundary does not have a finite mass, and hence is unphysical (and beyond a certain $\xi$ it would also be unstable \cite{A,LW,Pad,Katz}).  However, here we are postulating a perfectly reflecting sphere at some radius $r=R$ to confine the isothermal plasma and keep the total mass inside at the fixed value of the solar mass $M_\odot$. 

If we say that the value of $\xi = r/L$ at the surface (the reflecting sphere of radius $R$) is $X$ (henceforth not to be confused with the previous use of $X$ as the mass fraction of hydrogen in the solar composition), then, using the Emden-Chandrasekhar Eq.\ (\ref{EC}) and $\rho_c = kT/(4\pi G m L^2)$,
\ba
M(X) &=& \int_0^R 4\pi r^2 \rho dr = \int_0^X 4\pi L^3 \xi^2 d\xi\rho_c e^{-\psi} = 4\pi L^3 \rho_c \int_0^X \frac{d}{d\xi}\left(\xi^2\frac{d\psi}{d\xi}\right)d\xi\nonumber \\
&=& 4\pi L^3 \rho_c X v(X) = \frac{RkTv}{Gm}.
\label{M(X)}
\ea
Therefore, with $v$ the value of $v(\xi)$ at the surface, $\xi = X$,
the temperature of the isothermal plasma fluid ball of solar mass $M_\odot$ is
\bb
T = \frac{GM_\odot m}{kRv} = \left(\frac{R_\odot}{R}\right)\frac{2T_\odot}{v}.
\label{T(v)}
\ee

The Chandrasekhar-Wares \cite{CW} tabulated values of $\psi$, $e^{-\psi}$, $\psi' \equiv d\psi/d\xi$, $u \equiv \xi e^{-\psi}/\psi'$, $v \equiv \xi\psi'$, $\xi v = \xi^2\psi'$, and $\rho_c/\bar{\rho} = \xi^2/(3v) = \xi/(3\psi')$ show that $v$ has a maximum value near $\xi = 9$ of $v_\mathrm{max} \approx 2.5176$, so as the confined sun burns hydrogen to helium and increases $E$, which decreases $X$, the temperature $T$ initially drops until it reaches a minimum value of
\bb
T_\mathrm{min} = \left(\frac{R_\odot}{R}\right)\frac{2T_\odot}{v_\mathrm{max}}
\approx (5.665\times 10^6\ \mathrm{K})\left(\frac{R_\odot}{R}\right).
\label{Tmin}
\ee
when $X = R/L$, the surface value of $\xi = r/L$, drops to approximately 9.
At this minimum in the temperature, the ratio of the central density to the average density is
\bb
\frac{\rho_c}{\bar{\rho}} = \frac{\xi^2}{3v} \approx 10.71
\label{densityconcentration}
\ee
Since the average density of a solar mass $M_\odot$  in a sphere of radius $R$ is
\bb
\bar{\rho} = \frac{3M_\odot}{4\pi R^3} = \rho_\odot\left(\frac{R_\odot}{R}\right)^3 \approx (1\,410\ \mathrm{kg/m}^3)\left(\frac{R_\odot}{R}\right)^3,
\label{rhohat}
\ee
the central density at the temperature minimum will be about 10.71 times this, or
\bb
\rho_c \approx (15\,100\ \mathrm{kg/m}^3)\left(\frac{R_\odot}{R}\right)^3.
\label{rhocentral}
\ee

The kinetic energy of the $N_\odot = M_\odot/m$ nonrelativistic particles in the 
isothermal plasma ball of the mass and composition of the sun and inside a perfectly reflecting shell of radius $R$ and with $v = -(r/\rho)d\rho/dr$ at this shell at the outer surface of the ball is
\bb
K = \frac{3}{2}N_\odot kT = \frac{3GM_\odot^2}{2Rv} 
= E_\odot \left(\frac{R_\odot}{R}\right)\frac{3}{v},
\label{K}
\ee
The gravitational potential energy is, with $M(r)$ being the mass inside radius $r = L\xi$,
\ba
U &=& -\int_0^R \frac{GM(r)}{r}\frac{dM(r)}{dr}dr = -\frac{L}{G}\left(\frac{kT}{m}\right)\int_0^X uv^2 d\xi = -\frac{L}{G}\left(\frac{GM_\odot}{Rv}\right)^2 X v(3-u)
\nonumber \\
&=& -\frac{GM_\odot^2}{R}\left(\frac{3-u}{v}\right) = E_\odot\left(\frac{R_\odot}{R}\right)\left(\frac{2u-6}{v}\right).
\label{U}
\ea
Here the integral can be evaluated by using the fact that
\bb
\frac{d}{d\xi}[\xi v(3-u)] = uv^2,
\label{dI}
\ee
which one can easily see follows from Eqs. (\ref{du}) and (\ref{dv}).

One also has the energy density of the electromagnetic radiation, which is
\bb
E_r = \frac{4\pi}{3}R^3 a T^4 = \frac{R^3(kT)^4}{45(hc)^3} = E_0\left(\frac{R_\odot}{R}\right)\frac{\lambda}{v^4}
\label{Er}
\ee
with
\bb
\lambda = \frac{8\pi^3}{45N_\odot^4}\left(\frac{GM_\odot^2}{\hbar c}\right)^3 = \frac{8\pi^3}{45}\frac{M_\odot^2 m^4}{m_{\mathrm{Pl}}^6} \approx 0.2328,
\label{lambda}
\ee
where
\bb
m_{\mathrm{Pl}} \equiv \sqrt{\frac{\hbar c}{G}} = \sqrt{\frac{h c}{2\pi G}} = 2.176\,434(24)\times 10^{-8}\ \mathrm{kg}
\label{mPl}
\ee
is the Planck mass.

Therefore, the total energy of the isothermal plasma fluid ball, not counting the rest mass energies of the electrons and nuclei, is
\bb
E = K + U + E_r = E_0\left(\frac{R_\odot}{R}\right)\left(\frac{2u-3}{v}+\frac{\lambda}{v^4}\right) = E_0\left(\frac{R_\odot}{R}\right)P,
\label{E(xi)}
\ee
where
\bb
P = \frac{2u-3}{v}+\frac{\lambda}{v^4}
\label{P}
\ee
just depends on $X = \xi = R/L$ at the perfectly reflecting sphere at the surface of the isothermal ball of plasma, through the solution $\psi(\xi)$ of the Emden-Chandrasekhar Eq.\ (\ref{EC}) and the auxiliary dimensionless functions $u = \xi e^{-\psi}/(d\psi/d\xi)$ and $v = \xi d\psi/d\xi$ evaluated at the surface.

Making a quadratic interpolation of the Chandrasekhar-Wares tabulated values \cite{CW} for $u$ and $v$ for $\xi = 32$, $\xi = 34$, and $\xi = 36$ to calculate $P(\xi)$ for intermediate values shows that the minimum value for $P(\xi)$ is
\bb
P_m \approx -0.6556
\label{Pm}
\ee
at $\xi \approx 33.36$, giving the minimum energy for an isothermal plasma ball of the solar mass and composition and radius $R$ as
\bb
E_m \approx -0.6556\, E_0\left(\frac{R_\odot}{R}\right) \approx -(1.243\times 10^{41}\ \mathrm{J})\left(\frac{R_\odot}{R}\right).
\label{Em}
\ee

Because of the positive contribution of the thermal radiation, this minimum value for $P$ is slightly less negative than the minimum value without including the thermal radiation, which Padmanabhan \cite{Pad} calculated as approximately $-0.670.$

In contrast, the work required to dissipate the present solar matter to infinity is about $6.6\times 10^{41}$ J \cite{Allen}, which if one sets $R = R_\odot$ corresponds to $P \approx -3.48$.  Therefore, the present sun is too tightly bound gravitationally to be able to form an isothermal ball of plasma, which if enclosed by a perfectly reflecting sphere of radius $R = R_\odot$ would require the sun to generate additional energy
\bb
\Delta E \approx 2.8 E_\odot \approx 5.4\times 10^{41}\ \mathrm{J}
\label{DeltaE}
\ee
which at the present solar luminosity would require about 44 million years.  However, since this time is short compared with the solar lifetime of the present sun and with the lifetime for the sun to be confined inside a perfectly reflecting sphere of radius $R \geq R_\odot$, I shall ignore the time needed for the sun to produce enough energy to become an isothermal sphere at the minimum energy $E_m$ given above.  Huston and Wright \cite{H,HW} raise the possibility that even after reaching the energy of an isothermal sphere, thermalization might take a long time for stars with radiative exteriors such as the sun, and I am also ignoring that time which I have not calculated.

\section{Formulas for the Isothermal Hydrogen Burning Time Inside a Perfectly Reflecting Sphere}

Now I shall estimate the lifetime of the sun once it has been put inside a perfectly reflecting sphere of radius $R$ and has reached an isothermal configuration at the minimum value of $P$, $P_m \approx -0.6556$.  I shall assume that most of the time will be spent burning hydrogen to helium by the PPI chain, and I shall use Eq.\ (5-30) on page 378 of Clayton \cite{Clayton} to give the rate.  In cgs units with hydrogen mass fraction $X_H$ and $T_6$ being the temperature in units of $10^6$ K, the energy production rate per time and per mass is given as
\ba
\epsilon_\mathrm{PPI} = 2.32\times 10^6 \rho X_H^2 T_6^{-\frac{2}{3}}
\exp{(-33.81 T_6^{-\frac{1}{3}})}(1+0.0123T_6^{\frac{1}{3}}\!\!\!\!&+&\!\!\!\!0.0109T_6^{\frac{2}{3}}+0.00095T_6)\nonumber \\ 
&&\mathrm{erg\ g^{-1}\ sec^{-1}}.
\label{epsiloncgs}
\ea

Most of the time needed for the hydrogen burning inside the perfectly reflecting sphere will occur when $T_6$ is not large enough for the series in $T_6^{\frac{1}{3}}$ to be much larger than 1.  In particular, the minimum temperature when $R = R_\odot$ from Eq.\ (\ref{Tmin}) gives $T_6 \approx 5.665$ and $T_6^{1/3} \approx 1.783$, which makes the series have the value
\bb
1+0.0123T_6^{\frac{1}{3}} + 0.0109T_6^{\frac{2}{3}} + 0.00095T_6 \approx 1.0619,
\label{series}
\ee
which is within about 6\% of unity.  Of course, for larger $R/R_\odot$ and hence smaller minimum $T_6 \approx 5.665(R_\odot/R)$, the series will be even closer to unity.  Therefore, I shall neglect all but the first term in the series in $T_6^{\frac{1}{3}}$.  Then setting the hydrogen fraction by mass at $X_H = 0.70$ as I have done, and expressing quantities in terms of the solar parameters rather than in terms of cgs units, the energy generation rate, the rate for converting part of the rest mass energy of the nuclei to $E = K + U + E_r$ (kinetic, gravitational potential, and radiation energies) is
\bb
\frac{dE}{dt} = \int\epsilon_{PPI}dM = \frac{\epsilon_\odot}{\rho_\odot}F(\tau)\int\rho dM = \frac{\epsilon_\odot}{\rho_\odot}F(\tau) J,
\label{dE/dt}
\ee
where
\bb
\tau \equiv \frac{T}{2T_\odot} = \frac{R_\odot}{Rv},
\label{tau}
\ee
is a dimensionless temperature that for an isothermal fluid ball varies inversely with the radius $R$ of the reflecting sphere at the surface and also inversely with the function $v(\xi) = \xi d\psi/d\xi = -d\ln{\rho}/d\ln{r}$ evaluated at the surface of the ball, and Eq.\ (\ref{epsiloncgs}) with the series truncated to the first term (1) leads to
\bb
F(\tau) \approx 142\,000\, \tau^{-2/3} \exp{(-13.94 \tau^{-1/3})}.
\label{F(tau)}
\ee

I am also defining the dimensionful
\bb
J \equiv \int \rho dM = 4\pi L^3 \rho_c^2 \int_0^X\xi^2 e^{-2\psi}d\xi 
= \frac{M_\odot^2 X j}{4\pi R^3 v^2},
\label{J}
\ee
where $X$ and $v$ are the values of $\xi$ and of $v(\xi)$ at the surface, and where the dimensionless analogue of $J$ is
\bb
j(X) \equiv \int_0^X \xi^2 e^{-2\psi} d\xi.
\label{j}
\ee

I could not find any explicit exact formula for $j(X)$ in terms of $X$, $u(X)$, and $v(X)$, but to avoid doing a numerical integration for it, I used the approximation \cite{PE} that
\bb
e^{-\psi} \approx e^{-\psi_{PE}} = \frac{50}{10+\xi^2}-\frac{48}{12+\xi^2},
\label{psiPE}
\ee
which leads to the following explicit approximation for $j(X)$:
\bb
j(X) \approx j_{PE}(X) = 2525\sqrt{10}\tan^{-1}{\frac{X}{\sqrt{10}}} 
- \frac{1250X}{10+X^2}
-4608\sqrt{3}\tan^{-1}{\frac{X}{\sqrt{12}}}
- \frac{1152 X}{12+X^2}.
\label{jPE}
\ee
Estimates showed that up to $X \approx 33.36$, where the energy is minimized, $j_{PE}(X)$ is only a few percent larger than $j(X)$, and the error of using it rather than $j(X)$ has the opposite sign as the error of dropping the higher powers of $T_6^{1/3}$ in the series in Eq.\ (\ref{epsiloncgs}), with both errors small and of the same general order of magnitude.

Now, after some algebra, one can show that the approximate formulas above lead to the time (in units of the characteristic solar time $t_\odot \equiv E_\odot/L_\odot \equiv (GM_\odot^2)/(2R_\odot L_\odot) \approx 1.570\times 10^7\ \mathrm{years}$) for the hydrogen burning stage (which is expected to take up the majority of the lifetime of sun when enclosed in a perfectly reflecting sphere of radius $R$, at least if the thermalization time is not enormously larger than $t_\odot$) to be
\bb
\frac{\Delta t}{t_\odot} \approx \alpha(R/R_\odot)^{4/3}\int_{34}^1\frac{v^{4/3}}{X j_{PE}(X)}[\exp{(\beta v^{1/3})}]^{(R/R_\odot)^{1/3}} \frac{dP}{dX} dX
\label{Delta t/tS}
\ee
where the numerical constants are $\alpha \approx 2\times 10^{-5}$ and $\beta \approx 13.94$, $j_{PE}(X)$ is given by Eq.\ (\ref{jPE}), and $P = (2u-3)/v + \lambda/v^4$, as given by Eq.\ (\ref{P}).  Remember that $X = R/L$ is the value of the dimensionless radial variable $\xi = r/L$ at the surface of the isothermal plasma ball (the location of the perfectly reflecting shell), and $v$ is the value of $\xi d\psi/d\xi = - d\ln{\rho}/d\ln{r}$ at the surface.  I am taking the integration in the direction of the time evolution, from $X = 34$ that is near where the energy $E$ is the minimum for an isothermal fluid ball (ignoring the relatively short time of the evolution for enough energy to be emitted to get to this point and to thermalize), down to a small value of $X$, chosen here to be 1, beyond which the evolution would become much more rapid because of the neglected terms in $\epsilon_{PPI}$ and because of helium burning.  (The value of the integral depends only weakly on where this cutoff is located, especially for $R$ several times $R_\odot$.)  In the direction of the time evolution, $P$ increases, as both $dP/dX$ and $dX$ are negative.

\section{Results of the Numerical Integrations}

Rather than spend time programming a computer to do the numerical integrations, and also to highlight the fact that the results can fairly easily be obtained from simple calculations using published results such as Clayton's formula for the PPI chain hydrogen burning energy production rate \cite{Clayton} and the Chandrasekhar-Wares tables \cite{CW} of the solution of the Emden-Chandrasekhar equation for the structure of a self-gravitating isothermal perfect fluid ball, I have done the numerical integrations of Eq.\ (\ref{Delta t/tS}) mainly by hand with an HP 35s Scientific Calculator, using its programmable features only to calculate the functions in that equation, including $P(X)$ and $j_{PE}(X)$ for integer values of $X$ ($\xi$ at the plasma fluid ball surface where it is contained by a perfectly reflecting spherical shell) between 1 and 34 (except for 31 and 33, which were not given in the tables).  I did the calculations for $R/R_\odot =$ 1, 2, 3, 4, 5, 10, 20, 40, 83.24 (the semimajor axis of the orbit of Mercury), 155.54 (the semimajor axis of the orbit of Venus), and 215.03 (one astronomical unit, essentially the semimajor axis of the orbit of the earth).

For each step, I used the trapezoid rule by multiplying the change in $P$ (the integral of $(dP/dX)dX$ from the beginning to the end of the step) by the average of the rest of the integrand in Eq.\ (\ref{Delta t/tS}) at the two endpoints and added the result for all steps.  For $R/R_\odot =$ 1, 2, and 3, the sum of the contributions for the steps between $X=16$ and $X=34$ were less than 1.7\% of the total, so for larger values of $R/R_\odot$ I just calculated the 15 steps between $X=1$ and $X=16$ and approximated the uncalculated steps by a geometric series matched to the last two calculated steps, which also always contributed less than 2\% to the total.

I then tried various fitting functions for the resulting estimates of the lifetimes of the hydrogen burning phases while the solar material was an isothermal plasma ball of radii given by the 11 values of $R$ above.  For $R/R_\odot \leq 5$, I found an excellent fit was given by the following 2-parameter simple exponential function
\bb
\Delta t_2 = (5.4\ \mathrm{yr})\exp{[20.825(R/R_\odot)^{1/3}]},
\label{Delta t_2}
\ee
which gave a relative discrepancy of less than 0.5\% for all five values of $R/R_\odot =$ 1, 2, 3, 4, and 5.  (The exponent 1/3 of the $(R/R_\odot)$ inside the exponential came from the $T_6^{1/3}$ factor in the exponential in the Clayton formula Eq.\ (\ref{epsiloncgs}) for the PPI chain energy production rate, so it was not one of the 2 parameters I used to fit the results.)

For fitting all 11 values, I did not find any simple 2-parameter formula that fit well over the whole range of $R/R_\odot$ from 1 to 215.03, but after trying about five different forms, I found the following function with 3 free parameters that I fit to the numerical results for $R/R_\odot =$ 1, 10, and 215.03 to match all 11 numerical results to within 9\%, which is itself roughly my crude estimate of the error in the calculated lifetimes inside the 11 different values of $R/R_\odot$:
\bb
\Delta t_3 = (50\ \mathrm{yr})\exp{[18.6(R/R_\odot)^{1/3}]}
\left(\frac{R}{R_\odot}\right)^{\frac{4}{3}\{1+0.2\ln{[(R/R_\odot)^{1/3}/6]}\}}.
\label{Delta t_3}
\ee
Here the 3rd fitting parameter (besides the 50 and the 18.6 that are analogous to the 5.4 and 20.825 in $\Delta t_2$) is the 0.2 coefficient of the logarithm in the exponent of $R/R_\odot$; the denominator of 6 in the argument of that logarithm was chosen so that the logarithm would be nearly 0 when the perfectly reflecting sphere is at one astronomical radius, $R \approx 215.03 \sim 6^3$, to simplify the fitting procedure.

Table~\ref{table} gives the calculated lifetimes for hydrogen burning for the solar composition confined inside perfectly reflecting spheres of various radii $R$, in units of the present solar radius $R_\odot$, along with the values given by the 2-parameter and 3-parameter approximate fitting formulas $\Delta t_2$ and $\Delta t_3$ of Eqs.\ (\ref{Delta t_2}) and (\ref{Delta t_3}) respectively, as well as the ratios of those approximations to the numerical calculations of the lifetimes.  Note that the 2-parameter fitting formula $\Delta t_2$ works better for $1 \leq R/R_\odot \leq 5$, but the 3-parameter fitting formula $\Delta t_3$ works better overall for $1 \leq R/R_\odot \leq 215.03$.

\begin{figure}[h]
    \centering
    \includegraphics[width=0.8\textwidth]{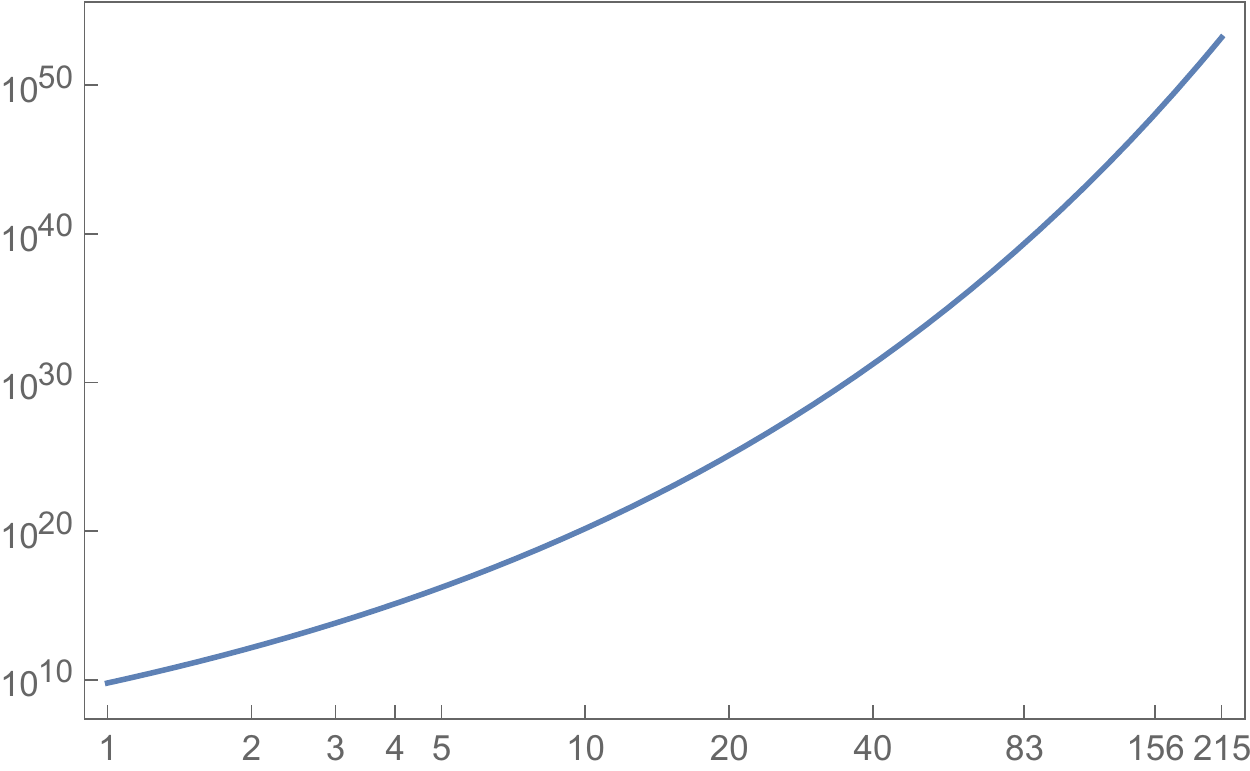}
    \caption{The confined solar lifetime in years versus the radius of the perfectly reflecting sphere at the surface in units of the present solar radius, $R/R_\odot$.}
    \label{fig}
\end{figure}

Figure~\ref{fig} gives a log-log plot the fitting function $\Delta t_3$ of Eq.\ (\ref{Delta t_3}), in years, as a function of $R/R_\odot$, the radius of the confining perfectly reflecting sphere at the surface divided by the current solar radius.  The horizontal axis is marked at the values of $R/R_\odot$ that were used in the simple numerical calculations done on a pocket calculator, using the PPI energy production rate formula from \cite{Clayton} and the numerical tables \cite{CW} of the Emden-Chandrasekhar isothermal function \cite{E,C}.

\begin{table}[h]
\begin{tabular}{|r|c|c|r|c|c|} \hline
$R/R_\odot$ & $\Delta t$ & $\Delta t_2$ & $\Delta t_2/\Delta t$ & $\Delta t_3$ & $\Delta t_3/\Delta t$ \\ \hline
1.00 & $5.90\times 10^9\;$ yr & $5.92\times 10^9\;$ yr & 1.003 & $5.98\times 10^9\;$ yr & 1.014 \\ \hline
2.00 & $1.31\times 10^{12}$ yr & $1.32\times 10^{12}$ yr & 1.008 & $1.42\times 10^{12}$ yr & 1.081\\
\hline
3.00 & $5.90\times 10^{13}$ yr & $5.89\times 10^{13}$ yr & 0.998 & $6.37\times 10^{13}$ yr & 1.079\\
\hline
4.00 & $1.21\times 10^{15}$ yr & $1.21\times 10^{15}$ yr & 0.995 & $1.29\times 10^{15}$ yr & 1.064\\
\hline
5.00 & $1.55\times 10^{16}$ yr & $1.55\times 10^{16}$ yr & 1.003 & $1.62\times 10^{16}$ yr & 1.049\\
\hline
10.00 & $1.46\times 10^{20}$ yr & $1.61\times 10^{20}$ yr & 1.108 & $1.45\times 10^{20}$ yr & 0.997\\
\hline
20.00 & $1.25\times 10^{25}$ yr & $1.86\times 10^{25}$ yr & 1.493 & $1.22\times 10^{25}$ yr & 0.975\\
\hline
40.00 & $1.77\times 10^{31}$ yr & $4.45\times 10^{31}$ yr & 2.517 & $1.66\times 10^{31}$ yr & 0.941\\
\hline
83.24 & $2.52\times 10^{39}$ yr & $1.60\times 10^{40}$ yr & 6.341 & $2.33\times 10^{39}$ yr & 0.925\\
\hline
155.54 & $1.09\times 10^{48}$ yr & $2.23\times 10^{49}$ yr & 20.390 & $9.99\times 10^{47}$ yr & 0.914\\
\hline
215.03 & $1.66\times 10^{53}$ yr & $7.76\times 10^{54}$ yr & 46.669 & $1.59\times 10^{53}$ yr & 0.958\\
\hline
\end{tabular}
\caption{Lifetime $\Delta t$ of the sun confined to the interior of a perfectly reflecting sphere of radius $R$, along with fitting formulas $\Delta t_2 = (5.4\ \mathrm{yr})\exp{[20.825(R/R_\odot)^{1/3}]}$ and $\Delta t_3 = (50\ \mathrm{yr})\exp{[18.6(R/R_\odot)^{1/3}]}
(R/R_\odot)^{\frac{4}{3}\{1+0.2\ln{[(R/R_\odot)^{1/3}/6]}\}}$.  Note that $R = 83.24R_\odot$ is the semimajor axis of the orbit of Mercury, $R = 155.54R_\odot$ is that of Venus, and $R = 215.03R_\odot$ is that of Earth.}
\label{table}
\end{table}

\section{Discussion}

Enclosing the sun inside a perfectly reflecting shell of the same radius as that of the present sun, $R=R_\odot$, does not increase the lifetime significantly.  (And in this case the corrections from the series in $T_6^{1/3}$ are not negligible, so I would expect the lifetime to be significantly less than my estimate.)  However, enclosing the sun inside a sphere of radius just twice as large, $R = 2R_\odot$, would increase the lifetime by several orders of magnitudes (even if in this case the corrections might be rather more than 10\%).  Making the reflecting shell radius even larger increases the lifetime greatly.  By the time one gets $R$ up to the orbital radii of the inner planets, the calculated lifetime is so long that it is likely to be invalidated by  baryon decay, which would seem to put a very firm upper limit on the lifetime of the sun that cannot be circumvented just by putting it into a perfectly reflecting shell.

Therefore, it seems that by putting the sun inside a perfectly reflecting shell of a suitable radius between the present radius of the sun and that of the orbits of the inner planets, one can get any lifetime between that of the unconfined sun and that of baryon decay. 

It is tempting to speculate whether some very advanced civilization might be able to circumvent the extremely severe practical constraints I have ignored here and construct such a shell or Dyson sphere around a star to extend the stellar lifetime as long as desired, up to the time of baryon decay.  In principle one could construct an aperture in the shell to withdraw as much energy as the civilization needs without significantly reducing the lifetime (unless the civilization requires far more power than ours does, as it might if it needs to construct and maintain the Dyson sphere).  However, I shall leave it as a challenge for the future for how even in principle an advanced civilization might be enabled to survive past the time of baryon decay.

After finishing all the calculations and the bulk of the writing of this paper, I found that J.\ D.\ Bernal \cite{Bernal} had written a foretaste of what I have calculated could be the case if stars were surrounded by perfectly reflecting spheres to extend their lifetimes by millions of millions of times:

``A star is essentially an immense reservoir of energy which is being dissipated as rapidly as its bulk will allow. It may be that, in the future, man will have no use for energy and be indifferent to stars except as spectacles, but if (and this seems more probable) energy is still needed, the stars cannot be allowed to continue to in their old way, but will be turned into efficient heat engines. The second law of thermodynamics, as Jeans delights in pointing out to us, will ultimately bring this universe to an inglorious close, may perhaps always remain the final factor. But by intelligent organization the life of the universe could probably be prolonged to many millions of millions of times what it would be without organization. Besides, we are still too close to the birth of the universe to be certain about its death.'' 

\section{Acknowledgments}

This research was funded by the Natural Sciences and Engineering Research Council of Canada.  After I shared a nearly-final version of this paper with Macy Huston and Jason Wright, Jason Wright raised several helpful questions, some of which (such as whether a star such as the sun with radiative energy transfer can thermalize inside a perfectly reflecting shell before hydrogen burning is completed) will have to await future research.

\end{document}